\documentclass[10pt,preprint]{sigplanconf}

\usepackage[T1]{fontenc}
\usepackage{amsmath}
\usepackage{listings}
\usepackage{graphics}
\usepackage{epsfig}
\usepackage{bbding}
\usepackage{color}
\usepackage{url}
\usepackage{xspace,amsthm,amsmath,amssymb,url,microtype,amsfonts,color}
\usepackage{graphicx}
\usepackage{caption}
\usepackage{subcaption}
\usepackage{amscd,hyperref}
\usepackage{enumitem}
\usepackage{lstlang0}
\usepackage{fancyvrb}
\usepackage{pifont}
\usepackage{pgfplotstable}
\usepackage{colortbl}
\usepackage{hhline}
\usepackage{clrscode3e}

\newcommand{\cmark}{\ding{51}}

\newcommand{\surveyman}{\textsc{SurveyMan}\xspace}
\newcommand{\punt}[1]{}

\begin{document}

\conferenceinfo{UMass CS Technical Report UM-CS-2014-002}{}


\title{\surveyman{}: Programming and Automatically Debugging Surveys}


\authorinfo{Emma Tosch \and Emery D. Berger}
           {School of Computer Science \\
            University of Massachusetts, Amherst \\
            Amherst, MA  01003}
          {\{etosch,emery\}@cs.umass.edu}

\maketitle

\begin{abstract}
Surveys can be viewed as programs, complete with logic, control flow, and
bugs. Word choice or the order in which questions are asked can
unintentionally bias responses. Vague, confusing, or intrusive
questions can cause respondents to abandon a survey. Surveys can also
have runtime errors: inattentive respondents can taint results. This
effect is especially problematic when deploying surveys in
uncontrolled settings, such as on the web or via crowdsourcing
platforms. Because the results of surveys drive business decisions and
inform scientific conclusions, it is crucial to make sure they are
correct.

We present \surveyman{}, a system for designing, deploying, and
automatically debugging surveys. Survey authors write their surveys in a lightweight
domain-specific language aimed at end users. \surveyman{}
statically analyzes the survey to provide feedback to survey authors
before deployment. It then compiles the survey into JavaScript and
deploys it either to the web or a crowdsourcing
platform. \surveyman{}'s dynamic analyses automatically find survey
bugs, and control for the quality of responses. We
evaluate \surveyman{}'s algorithms analytically and empirically,
demonstrating its effectiveness with case studies of social science
surveys conducted via Amazon's Mechanical Turk.

\end{abstract}

\section{Introduction}
Surveys and polls are widely used to conduct research for industry,
politics, and the social sciences. Businesses use surveys to perform
market research to inform their spending and product
strategies~\cite{churchill2009marketing}. Political and news
organizations use surveys to gather public opinion, which can
influence political decisions and political campaigns. A wide range of
social scientists, including psychologists, economists, health
professionals, political scientists, and sociologists, make extensive
use of surveys to drive their
research~\cite{berinsky2012evaluating,de2002surveys,horton2011online,mason2012conducting}.

In the past, surveys were traditionally administered via mailings,
phone calls, or face-to-face interviews~\cite{dillman1978mail}. Over
the last decade, web-based surveys have become increasingly popular as
they make it possible to reach large and diverse populations at low
cost~\cite{cogprints2357,tourangeau2013science,Couper:2008:DEW:1457607,umbach2004web}.
Crowdsourcing platforms like Amazon's Mechanical Turk make it possible
for researchers to post surveys and recruit participants at scales
that would otherwise be out of reach.

Unfortunately, the design and
deployment of surveys can seriously threaten the validity of their
results:

\paragraph{Question order effects.}
Placing one question before another can lead to different
responses than when their order is reversed. For example, a Pew
Research poll found that people were more likely to favor civil
unions for gays when this question was asked after one about whether
they favored or opposed gay marriage~\cite{pew-order}.

\paragraph{Question wording effects.}
Different question variants can inadvertently elicit wildly different
responses. For example, in 2003, Pew Research found that American
support for possible U.S. military action in Iraq was 68\%, but this
support dropped to 43\% when the question mentioned possible American
casualties~\cite{pew-wording}. Even apparently equivalent questions
can yield quite different responses. In a 2005 survey, 51\% of
respondents favored ``making it legal for doctors to \emph{give terminally
ill patients the means to end their lives}'' but only 44\% favored
``making it legal for doctors to \emph{assist terminally ill patients in
committing suicide}.''~\cite{pew-wording}

\paragraph{Survey abandonment.}
Respondents often abandon a survey partway through, a phenomenon known
as \emph{breakoff}. This effect may be due to \emph{survey
fatigue}, when a survey is too long, or because a particular
question is ambiguous, lacks an appropriate response, or is too
intrusive. If an entire group of survey respondents abandon a survey,
the result can be \emph{selection bias}: the survey will exclude an
entire group from the population being surveyed. 

\paragraph{Inattentive or random respondents.}
Some respondents are inattentive and make arbitrary choices, rather
than answering the question carefully. While this problem can arise in
all survey scenarios, it is especially acute in an on-line setting
where there is no direct supervision. To make matters worse, there is
no ``right answer'' to check against for surveys, making controlling
for quality difficult.

Downs et al.\ found that nearly 40\% of survey respondents on Amazon's
  Mechanical Turk answered
  randomly~\cite{Downs:2010:YPG:1753326.1753688}. So-called \emph{attention
  check} questions aimed at screening inattentive workers are
  ineffective because they are easily recognized. An unfortunately
  typical situation is that described by a commenter on a recent
  article about taking surveys on Mechanical Turk:

\begin{quote}
If the requester is paying very little, I will go
as fast as I can through the survey making sure to pass their
attention checks, so that I'm compensated fairly. Conversely, if the
requester wants to pay a fair wage, I will take my time and give a
more thought out and non random response.~\cite{npr-surveys}
\end{quote}


\noindent
While all of the above problems are known to
practitioners~\cite{Martin2006}, there is currently no way to address
them automatically. The result is that current practice in deploying
surveys is generally limited to an initial pilot study followed by a
full deployment, with no way to control for the potentially
devastating impact of survey errors and inattentive respondents.

From our perspective,
this is like writing a program, making sure it compiles,
and then shipping it---to run on a system with hardware problems. 

\subsection{\surveyman{}}

In this paper, we adopt the view that surveys are effectively
programs, complete with logic, control flow, and bugs. We
describe \surveyman{}, which aims to provide a scientific footing for
the development of surveys and the analysis of their
results. Using \surveyman{}, survey authors use a lightweight
domain-specific language to create their surveys.
\surveyman{} then deploys their surveys
over the Internet, either by hosting it as a website, or via
crowdsourcing platforms.

The key idea behind \surveyman{} is that by giving survey authors a
way to write their surveys that steers them away from
unnecessary ordering constraints, we can apply static analysis,
randomization, and statistical dynamic analysis to locate survey errors
and ensure the quality of responses.

\begin{figure}[!t]
 \centering
 \includegraphics[width=0.5\textwidth]{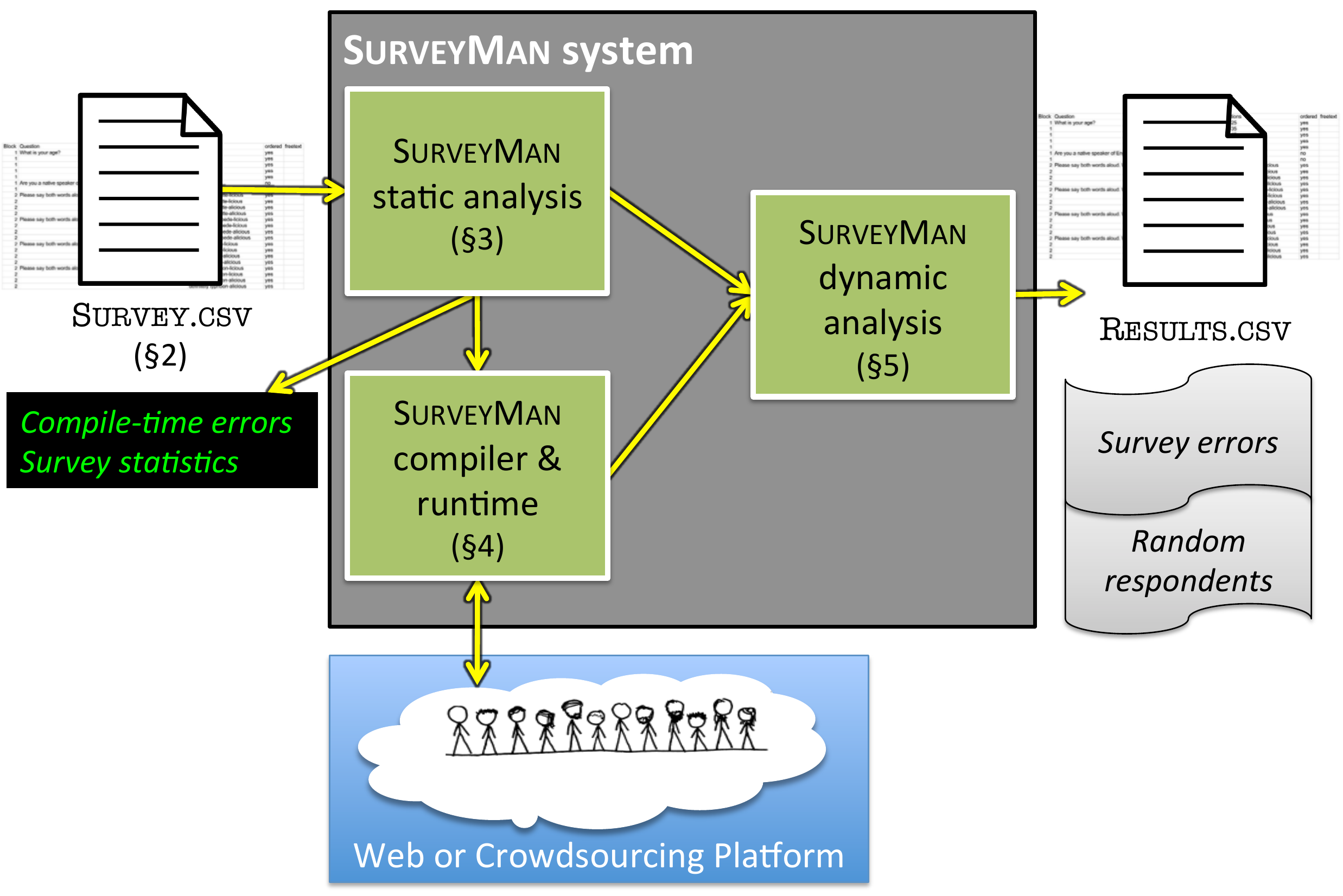}
 \caption{Overview of the \surveyman{} system.\label{fig:system-diagram}}
\end{figure}

\paragraph{Overview:} Figure~\ref{fig:system-diagram} depicts the \surveyman{}
workflow. Survey authors create surveys using the \surveyman{}
programming language. The \surveyman{} static analyzer checks
the \surveyman{} program for validity and reports key statistics about
the survey prior to deployment. If the program is
correct, \surveyman{}'s runtime system can then deploy the survey via
the web or a crowdsourcing platform: each respondent sees a
differently-randomized version. \surveyman{}'s dynamic analysis
operates on information from the static analysis and the results of
the survey to identify survey errors and inattentive
respondents. Table~\ref{tab:analyses} summarizes the analyses
that \surveyman{} performs.

\paragraph{Domain-Specific Language.}
We designed the \surveyman{} language in concert with social
scientists to ensure its usability and accessibility to
non-programmers  ($\S$\ref{sec:language}). The approach we take leverages the tools that our
target audiences use: because social scientists extensively use both
Excel and R, which both feature native support for comma-separated
value files, we adopt a tabular format that can be entered
directly in a spreadsheet and saved as a \texttt{.csv} file.

\surveyman{}'s domain-specific language is simple but captures most features
needed by survey authors, including a variety of answer types and
branching based on answers to particular questions. In addition, because \surveyman{}'s
error analysis depends on randomization, its language is designed to
maximize \surveyman{}'s freedom to randomize
question order, question variants, and answers.

A user writing a survey with \surveyman{} can optionally specify a partial order
over questions by grouping questions into \emph{blocks}, identified by
number. All questions within the same block may be asked in any order,
but must strictly precede the following block (i.e., all the questions
in block 1 precede those in block 2, and so on).

\begin{table*}[!t]
\centering
\begin{tabular}{>{\small}l>{\small}l>{\small}l}
\hline
\multicolumn{3}{c}{\emph{Static Analyses}} \\
\hline
\textbf{Well-formedness}   & $\S$\ref{sec:static-analysis-branches}  & Ensures survey is a DAG and other correctness checks \\
\textbf{Survey statistics} & $\S$\ref{sec:static-analysis-reports}   & Min, max, and avg. \# questions in survey; finds short-circuits and guides pricing  \\
\textbf{Entropy}           & $\S$\ref{sec:static-analysis-reports}   & Measures information content of survey; higher is better \\
\hline
\multicolumn{3}{c}{\emph{Dynamic Analyses}} \\
\hline
\textbf{Correlated Questions} & $\S$\ref{sec:analysis-correlated-questions} & Reports redundant questions which can be eliminated to reduce survey length \\
\textbf{Question Order Bias}  & $\S$\ref{sec:analysis-question-order-bias}  & Reports questions whose results depend \emph{on the order} in which they are asked \\
\textbf{Question Wording Variant Bias} & $\S$\ref{sec:analysis-wording-variant-bias} & Reports questions whose results depend \emph{on the way} they are worded \\
\textbf{Breakoff}                      & $\S$\ref{sec:analysis-breakoff} & Finds problematic questions that lead to survey abandonment \\
\textbf{Inattentive or Random Respondents} & $\S$\ref{sec:analysis-random-respondents} & Identifies unconscientious respondents so they can be excluded in analysis \\
\end{tabular}
\caption{The analyses that \surveyman{} performs on surveys and their deployment.\label{tab:analyses}}
\end{table*}

\paragraph{Static Analysis.}
\surveyman{} statically analyzes the
survey to verify that it meets certain criteria, such as that all
branches point to a target, and that there are no cycles ($\S$\ref{sec:static-analysis}). It
also provides additional feedback to the survey designer, indicating
potential problems with the survey that it locates prior to
deployment.

\paragraph{Compiler and Runtime System.}
\surveyman{} can then deploy the survey
either on the web with \surveyman{} acting as a webserver, or by
posting jobs to a crowdsourcing platform such as Amazon's Mechanical
Turk  ($\S$\ref{sec:runtime-system}). Each survey is delivered as a
JavaScript and JSON payload that manages presentation and flow through
the survey, and performs per-user randomization of question and answer
order subject to constraints placed by the survey author. When a
respondent completes or abandons a survey, \surveyman{} collects the
survey's results for analysis.

\paragraph{Dynamic Analyses.}
To find errors in a deployed survey, \surveyman{} performs statistical
analyses that take into account the survey's static control flow, the
partial order on questions, and collected results
($\S$\ref{sec:dynamic-analysis}). These analyses can identify a range
of possible survey errors, including question order bias, wording
variant bias, questions that lead to breakoff, and survey
fatigue. \surveyman{} reports each error with its corresponding
question, when applicable.  It also identifies inattentive and random
respondents. The survey author can then use the resulting report to
refine their survey and exclude random respondents.

Note that certain problems with surveys are beyond the scope of any
tool to address~\cite{umbach2004web}. These include \emph{coverage
error}, when the respondents do not include the target population of
interest; \emph{sampling error}, when the make-up of the respondents
is not representative of the target of interest;
and \emph{non-response bias}, when respondents to a survey are
different from those who choose not to or are unable to participate in
a survey. \surveyman{} can help survey designers limit non-response
bias due to abandonment by diagnosing and fixing its cause (breakoff or
fatigue).

\paragraph{Evaluation.}
Our collaborators in the social sciences developed a number of surveys
in \surveyman{}, which were deployed on Amazon's Mechanical
Turk ($\S$\ref{sec:evaluation}). We describe these experiences
with \surveyman{} and the results of the deployment, which identified
a number of errors in the surveys as well as random respondents.

\subsection{Contributions}

The contributions of this paper are the following:

\begin{itemize}

\item \textbf{Domain-Specific Language.} We introduce the \surveyman{}
domain-specific language for writing surveys, which enables error
detection and quality control by relaxing ordering constraints ($\S$\ref{sec:language},
$\S$\ref{sec:runtime-system}).

\item \textbf{Static Analyses.} We present static analyses for
identifying structural problems in surveys
($\S$\ref{sec:static-analysis}).

\item \textbf{Dynamic Analyses.} We present dynamic analyses to
identify a range of important survey errors, including question
variant bias, order bias, and breakoff, as well as inattentive or
random respondents ($\S$\ref{sec:dynamic-analysis}).

\item \textbf{Experimental Results.} We report on a deployment of \surveyman{}
with social scientists and demonstrate its utility
($\S$\ref{sec:evaluation}).

\end{itemize}

\section{\surveyman{} Domain-Specific Language}
\label{sec:language}
\begin{figure*}[!t]
\footnotesize
\centering
\begin{filecontents*}{example.csv}
BLOCK,QUESTION,OPTIONS,EXCLUSIVE,ORDERED,BRANCH
1,What is your gender?,Male,,,
1,,Female,,,
1,,Other,,,
1,What country do you live in?,United States,,,2
1,,India,,,3
1,,Other,,,3
2,What state do you live in?,Alabama,,TRUE,
2,,Alaska,,TRUE,
2,,Arizona,,TRUE,
2,,Arkansas,,TRUE,
2,,California,,TRUE,
3,How much time do you spend on Mechanical Turk?,Less than 1 hour per week.,,TRUE,
3,,1-2 hours per week.,,TRUE,
3,,2-4 hours per week.,,TRUE,
3,,4-8 hours per week.,,TRUE,
3,,8-20 hours per week.,,TRUE,
3,,20-40 hours per week.,,TRUE,
3,,More than 40 hours per week.,,TRUE,
3,Check all the reasons why you use Mechanical Turk.,Fruitful way to spend time.,FALSE,,
3,,Primary source of income.,FALSE,,
3,,Secondary source of income.,FALSE,,
3,,To kill time.,FALSE,,
3,,I find the tasks to be fun.,FALSE,,
3,,I am currently unemployed.,FALSE,,
\end{filecontents*}
\pgfplotstabletypeset[
col sep=comma,
    string type,
    columns/BLOCK/.style={column name=\textsc{Block}, column type={|c}},
    columns/QUESTION/.style={column name=\textsc{Question}, column type={|l}},
    columns/OPTIONS/.style={column name=\textsc{Options}, column type={|l}},
    columns/EXCLUSIVE/.style={column name=\textsc{Exclusive}, column type={|c}},
    columns/ORDERED/.style={column name=\textsc{Ordered}, column type={|c}},
    columns/BRANCH/.style={column name=\textsc{Branch}, column type={|c|}},
    every head row/.style={before row=\hline,after row=\hline},
    every last row/.style={after row=\hline},
    every row no 3/.style={before row=\hline},
    every row no 6/.style={before row=\hhline{|=|=|=|=|=|=|}\rowcolor[gray]{0.9}},
    every row no 11/.style={before row=\hhline{|=|=|=|=|=|=|}},
    every row no 18/.style={before row=\hline\rowcolor[gray]{0.9}},
    every even row/.style={before row={\rowcolor[gray]{0.9}}},
]{example.csv}
\caption{An example survey written using the \surveyman{} domain-specific language, adapted from Ipeirotis~\cite{ipeirotis2010demographics}. For clarity, a horizontal line separates each question, and a double horizontal line separates distinct \emph{blocks}, which optionally define a partial order: all questions in block $i$ appear in random order before questions in blocks $j > i$. When blocks are not specified, all questions may appear in a random order. This relaxed ordering enables \surveyman{}'s error analyses.\label{fig:example}}
\end{figure*}

\subsection{Overview}

The \surveyman{} programming language is a tabular, lightweight
language aimed at end users. In particular, it is designed to both make
it easy for survey authors without programming experience to create
simple surveys, and let more advanced users create sophisticated
surveys. Because of its tabular format, \surveyman{} users can enter
their surveys directly in a spreadsheet application, such as Microsoft
Excel. Unlike text editors or IDEs, spreadsheet applications are tools
that our target audience knows well. \surveyman{} can read in
\texttt{.csv} files, which it then checks for validity
($\S$\ref{sec:static-analysis}), compiles and deploys
($\S$\ref{sec:runtime-system}), and reports results, including errors
($\S$\ref{sec:dynamic-analysis}).

A key distinguishing feature of \surveyman{}'s language is its support
for randomization, including question order, question variants, and
answers. From \surveyman{}'s perspective, more randomization is
better, both because it makes error detection more effective and
because it provides experimental control for possible biases
($\S$\ref{sec:dynamic-analysis}). \surveyman{} is designed so that
survey authors must go out of their way to state when randomization is
\emph{not} to be used. This approach encourages authors to avoid
imposing unnecessary ordering constraints.

\paragraph{Basic Surveys:}
To create a basic survey, a survey author simply lists questions and
possible answers in a sequence of rows. When the survey is deployed,
all questions are presented in a random order, and the order of all
answers is also randomized.

\paragraph{Ordering Constraints:}
Survey authors can assign numbers to questions that they wish
to order; we use the terminology of the survey literature and call these numbers
\emph{blocks}. Multiple questions can have the same number, which
causes them to appear in the same block. Every question in the same block will be
presented in a random order to each respondent. All questions inside
blocks of a lower number will be presented before questions inside
higher-numbered blocks.

\surveyman{}'s block construct has additional features that can give
advanced survey authors finer-grained control over ordering
($\S$\ref{sec:advanced-blocks}).

\paragraph{Logic and Control Flow:}
\surveyman{} also includes both logic and control flow in the form of
branches depending on the survey taker's responses. Each answer can
contain a target branch (a block), which is taken if the survey taker
chooses that answer.


\punt{
The mix of control
flow, blocks, and randomization raises language design and
implementation challenges. The untrammeled use of branches could lead
to counterintuitive results. For instance, a respondent could
theoretically experience different paths through a survey (answering
different questions) simply as a result of randomizaton. By limiting
the expressiveness of branches, \surveyman{} rules out this
possibility.
}

\begin{table}[!t]
\centering
\begin{tabular}{>{\small}l|>{\small}l}
\textbf{Column} & \textbf{Description} \\
\hline
\textsc{Question}  & The text for a question \\
\textsc{Options}   & Answer choices, one per row \\
\hline
\textsc{Block}     & Numbers used to partially order questions \\
\textsc{Exclusive} & Only one choice allowed (default) \\
\textsc{Ordered}   & Present options in order \\
\textsc{Branch}    & For this response, go to this block \\
\textsc{Randomize} & Randomize option orders (default) \\
\textsc{Freetext}  & Allow text entry, optionally constrained \\
\textsc{Correlated} & Used to indicate questions are correlated \\
\end{tabular}
\caption{Columns in \surveyman{}. All except the first two (\textsc{Question} and \textsc{Options}) are optional.\label{tab:columns}}
\end{table}

\subsection{Syntax}

Figure~\ref{fig:example} presents a sample \surveyman{} program, a
survey from a Mechanical Turk demographic survey modified to
illustrate \surveyman{}'s features~\cite{ipeirotis2010demographics}.

Every \surveyman{} program contains a first row of column headers,
which indicate the contents of the following rows. The only mandatory
columns are \textsc{Question} and \textsc{Options}; all other columns
are optional. If a given column is not present, its default value is
used. The columns may appear in any order. 

\paragraph{Blocks.}
Each question can have an optional \textsc{Block} number associated
with it. Blocks establish a partial order. Questions with the same
block number may appear in any order, but must precede all questions
with a higher block number. If no block column is specified, all
questions are placed in the same block, meaning that they can appear
in any order.

\paragraph{Questions and Answers.}
The \textsc{Question} column contains the question text, which may
include HTML. Users may wish to specify multiple variants of the same question
to control for or detect question wording bias. To do this, users place
all of the variants in a particular block and have every question branch
to the same target.

The survey author specifies each question as a series of consecutive
rows. A row with the \textsc{Question} column filled in indicates the
end of the previous question and the start of a new one. All other
rows leave this column empty (as in Figure~\ref{fig:example}).

\textsc{Options} are the possible answers for a question. \surveyman{}
treats a row with an empty question text field as belonging to the
question above it. These may be rendered as radio buttons, checkboxes,
or freetext. If there is only one option listed and the cell is empty,
this question text is presented to the respondent as instructions.

\paragraph{Radio Button or Checkbox Questions.}
By default, all options are \textsc{Exclusive}: the respondent can
only choose one. These correspond to ``radio button'' questions. If
the user specifies a \textsc{Exclusive} column and fills it with
\texttt{false}, the respondent can choose multiple options, which are
displayed as ``checkbox'' questions.

\paragraph{Ordering.}
By default, options are unordered; this corresponds to nominal
data like a respondent's favorite food, where there is no ordering
relationship. Ordered options include so-called Likert scales, where
respondents rate their level of agreement with a statement; when the
options comprise a ranking (e.g., from 1 to 5); and when ordering is
necessary for navigation (e.g., for a long list of countries).

When options are unordered, they are presented to each respondent as one
of $m!$ possible permutations of the answers, where $m$ is the
number of options. To order the options, the user must fill in a
\textsc{Ordered} column with the value \textsc{True}. When they are
ordered, they can still be randomized: they are either
presented in the forward or backwards order. The user can only force the
answers to appear in exactly the given order by also including a
\textsc{Randomize} column and filling in the value \textsc{False}.

Unordered options eliminate option order bias, because the
position of each option is randomized. They also make inattentive
respondents who always click on a particular option choice
indistinguishable from random respondents. This property lets
\surveyman{}'s quality control algorithm simultaneously identify both
inattentive and random responses.

\begin{figure*}[!t]
 \centering
 \begin{subfigure}[b]{0.3\textwidth}
                \includegraphics[width=\textwidth]{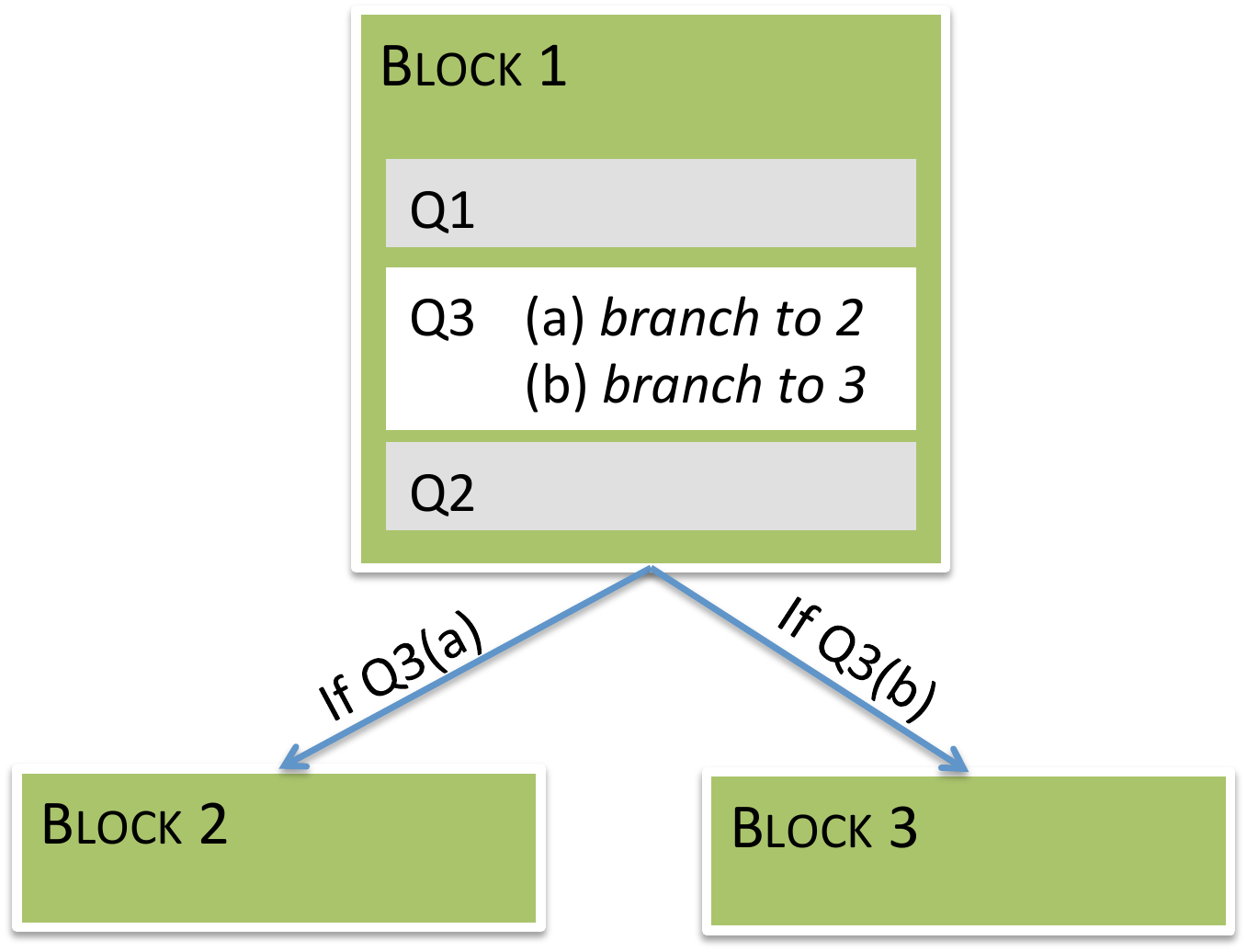}
                \caption{In this example of survey branches, the respondent is asked the branching question (Q3) as the second question. The execution of this branch is deferred until the respondent answers Q2, the last question in the block.}
                \label{fig:branch-diagram}
        \end{subfigure}%
 \quad\quad
 \begin{subfigure}[b]{0.3\textwidth}
                \includegraphics[width=\textwidth]{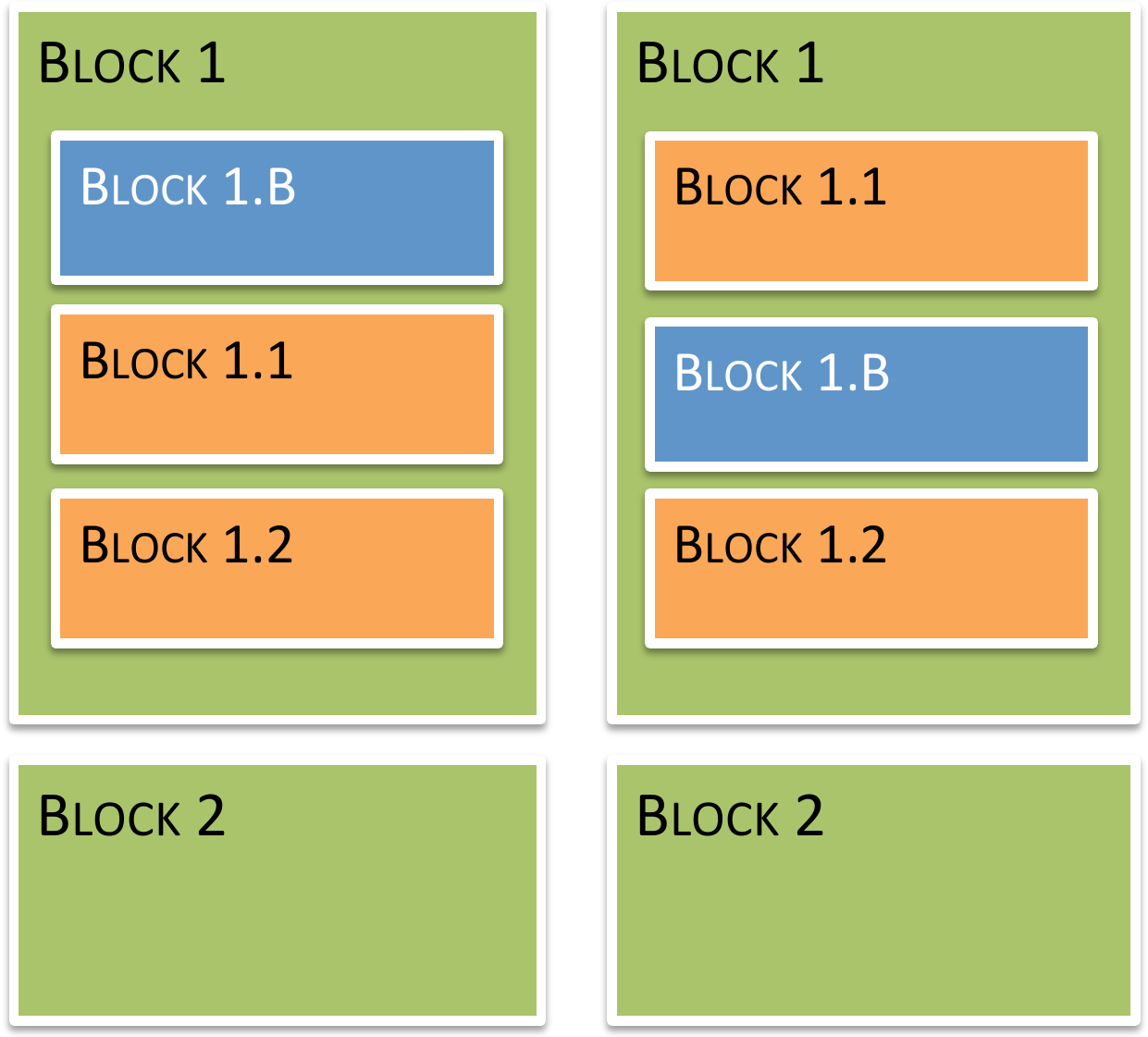}
                \caption{Two different instances of a survey with top-level blocks (\texttt{1} and \texttt{2}), nested blocks (\texttt{1.1} and \texttt{1.2}), and a floating block (\texttt{1.B}). The floating block can move anywhere within block \texttt{1}, while the other blocks retain their relative orders ($\S$\ref{sec:advanced-blocks}).}
                \label{fig:block-diagram}
        \end{subfigure}%
\caption{Examples of branches and blocks.\label{fig:blocks-and-branches}}
\end{figure*}

\paragraph{Branches.}
The \textsc{Branch} column provides control flow through the survey
when the survey designer wants respondents who answer a particular
question one way to take one path through the survey, while
respondents who answer differently take a different path. 
For example, in the survey shown in Figure~\ref{fig:example}, only
when respondents answer that they live in the United States will they
be asked what state they live in.

During the survey, branching is deferred until the end of a block, as
Figure~\ref{fig:blocks-and-branches}(\subref{fig:branch-diagram}) depicts. By avoiding premature branching, this
approach ensures that all users answer the same questions, regardless
of randomization. It also avoids the biasing of question order that
would result by forcing branches to appear in a fixed position.

Branching
from a particular question response must go to a higher numbered
block, preventing cycles.


\subsection{Advanced Features}

\surveyman{} several advanced features to give survey authors more control over their surveys and their interaction with \surveyman{}.

\paragraph{Correlated questions.}
\surveyman{} lets authors include a
\textsc{Correlated} column to assist in quality control. \surveyman{}
checks for correlations between questions to see if any redundancy can
be removed, since shorter surveys help reduce survey fatigue. However,
sometimes correlations are desired, whether to confirm a hypothesis or
as a form of quality control. \surveyman{} thus lets authors mark sets
of questions as correlated by filling in the column with arbitary
text: all questions with the same text are assumed to be
correlated. \surveyman{} will then only report if the answers to these
questions are \emph{not} correlated. If not, this information can be
used to help identify inattentive or adversarial respondents.

\subsubsection*{Advanced Blocks Notation}
\label{sec:advanced-blocks}

So far, we have described blocks as if they were limited to
non-negative integers. In fact, \surveyman{}'s block syntax is more
general\footnote{Formally, the syntax of blocks is given by the
  following regular expression:
  \texttt{[1-9][0-9]*(.[a-z1-9][0-9]*)*}.}.

There are three kinds of blocks: \emph{top-level blocks}, indicated by
a single number; \emph{nested blocks}, indicated by numbers separated
by periods; and \emph{floating blocks}, indicated by alphanumerics,
which can move around inside their parent blocks. We describe each in
terms of their interaction with other blocks and with
branches. Figure~\ref{fig:blocks-and-branches}(\subref{fig:block-diagram}) gives an example of each.

\paragraph{Top-level Blocks.}
A survey is composed of blocks. If there are no blocks (i.e., the
survey is completely flat), we say that all questions belong to a
single, top-level block (block \texttt{1}). Only one branch question
is allowed per top-level block. Additionally, branch question targets
must also be top-level blocks, though they cannot be floating blocks
(described below).

\paragraph{Nested Blocks.}
As with outlines, a period in a block indicates hierarchy: we can think of blocks
numbered \texttt{1.1} and \texttt{1.2} as being \emph{contained} within block
\texttt{1}. All questions numbered \texttt{1.1} precede all questions
numbered \texttt{1.2}, and both strictly precede blocks with numbers 2
or higher.

\paragraph{Floating Blocks.}
Survey authors may want certain questions to be allowed to appear anywhere
within a containing block. For example, a survey author might want
questions to appear within block \texttt{1}, but does not care whether
they appear before or after other questions contained in that block
(e.g., questions with block numbers \texttt{1.1} and
\texttt{1.2}). Such questions can be placed in \emph{floating
  blocks}. A question is marked as floating by using a
block name that is an alphanumeric string rather than a number.

In this case, the survey author could give a floating question the
block number \texttt{1.A}. That block is contained within block
\texttt{1}, and the question is free to float anywhere within that
block. The string itself has no effect on ordering; for example,
blocks \texttt{1.A} and \texttt{1.B} can appear in any order inside
block \texttt{1}. All questions in the same floating block float
together inside their containing block.

\punt{
\subsection{Semantics}

}

\section{Static Analyses}
\label{sec:static-analysis}

\surveyman{} provides a number of static analyses that check the
survey for validity and identify potential issues.

\subsection{Well-Formedness}
\label{sec:validity-checks}

After successfully parsing its \texttt{.csv} input, the \surveyman{}
analyzer verifies that the input
program itself is well-formed and issues warnings, if appropriate.
It checks that the header
contains the required \textsc{Question} and \textsc{Option} columns.
It warns the programmer if it finds any unrecognized headers, since these may
be due to typographical errors.
\surveyman{} also
issues a warning if it finds any duplicate questions.
It then performs more detailed analysis of branches:

\label{sec:static-analysis-branches}

\paragraph{Top-Level Branch Targets Only:}
\surveyman{} verifies that the target of each branch in the survey 
is a top-level block; that is, it cannot branch into a nested block. %
This check takes constant time.

\paragraph{Forward Branches:}
Since the survey must be a DAG, all branching must be to
higher-numbered blocks. \surveyman{} verifies that each branch target id
is larger than the id of the block that the branch question belongs to.

\paragraph{Consistent Block Branch Types:}
\surveyman{} computes the \emph{branch type} of
every block. A block can contain no branches in the block
(\textsc{none}) or exactly one branch question (\textsc{one}). It can
also be a block where all of the questions have branches to the same
destination (\textsc{all}). This is the approach used to express
question variants, where just one of a range of semantically-identical
questions is chosen. In this case, questions can only differ in their
text; only one of the variants (chosen randomly) will be asked.

\punt{
\begin{table*}[!t]
\centering
\begin{tabular}{>{\small}l|>{\small}l>{\small}l>{\small}l>{\small}l}
\textbf{Block Branch Type} & \textbf{Parent Type} & \textbf{Sibling Type} & \textbf{Children Type} & \textbf{Additional Sibling Constraints} \\
\hline
\textsc{none} & \{\textsc{none}, \textsc{one}\} & \{\textsc{none}, & \{\textsc{none}, \textsc{all}\}  &  If my parent is \textsc{one}, then only one of my siblings is \textsc{one}. \\
              &              & \textsc{one},                &          &  \\
              &              &  \textsc{all}\}             &          &  \\
\hline
\textsc{one}  & \{\textsc{one}\} &  \{\textsc{none}, & \{\textsc{none}, & If I immediately contain the branch question, \\
              &              &  \textsc{all}\} & \textsc{one},  &  then my children cannot be \textsc{one}. \\
              &              &                 & \textsc{all}\} &  Otherwise, exactly one of my descendants must be \textsc{one}. \\
\hline
\textsc{all}  & \{\textsc{none}, \textsc{one}\} & \{\textsc{none}, & \emph{N/A} &  If my parent is \textsc{one}, then only one of my siblings is \textsc{one}. \\ 
    &       &   \textsc{one}, &  &  I cannot have children. \\
    &       &   \textsc{all}\} & & \\ 
\end{tabular}
\caption{Block branch type constraints enforced by \surveyman{} (Section~\ref{sec:static-analysis-branches}).\label{tab:features}}
\end{table*}
}


\subsection{Survey Statistics and Entropy}
\label{sec:static-analysis-reports}

If a program passes all of the above checks, \surveyman{} produces a
report of various statistics about the survey, including an analysis
of paths through the survey.

If we view a path through the survey as the series of questions seen
and answered, the computation of paths would be intractable. Consider
a survey with 16 randomizable blocks, each having four variants and no
branching. We thus have $16!$ total block orderings; since we
choose from four different questions, we end up with $216!$ unique
paths.

Fortunately, the statistics we need require only computations of
the \emph{length} of paths rather than the \emph{contents} of the
path. The number of questions in each block can be computed statically. We
thus only need to consider the branching between these blocks
when computing path lengths.

\paragraph{Minimum Path Length:}
Branching in surveys is typically related to some kind of division in
the underlying population. When surveys have sufficient branching, it
may be possible for some respondents to answer far fewer questions
than the survey designer intended -- they may \emph{short circuit} the
survey. Sometimes this is by design: for a survey only interested in
curly-haired respondents, the survey can be designed so that answering
``no'' to ``Do you have curly hair?'' sends the respondent straight to
the end. In other cases, this may not be the intended effect and could
be either a typographical error or a case of poor survey
design. \surveyman{} reports the minimum number of questions a survey
respondent could answer while completing the survey.

\paragraph{Maximum Path Length:}
A survey that is too long can lead to survey fatigue. Surveys that are
too long are also likelier to lead to inattentive
responses. \surveyman{} reports the number of questions in the longest
path through the survey to alert authors to this risk.

\paragraph{Average Path Length:}
Backends such as Amazon Mechanical Turk require the requester to
provide a time limit on surveys and a payment. The survey author can
use the average path length through the survey to estimate the time it
would take to complete, and from that compute the baseline
payment. \surveyman{} computes this average by generating a large
number of random responses to the survey (currently, 5000) and reports
the mean length.

\paragraph{Maximum Entropy:}
Finally, \surveyman{} reports the entropy of the survey. This number
roughly corresponds to the complexity of the survey. If the survey has
very low entropy (few questions or answers), the survey will require
many respondents for \surveyman{} to be able to identify inattentive
respondents. Surveys with higher entropy provide \surveyman{} with
greater discriminatory power. For a survey of $n$ questions each with
$m_i$ responses, \surveyman{} computes a conservative upper bound on
the entropy on the survey as $n \log_2 (\max \lbrace m_i | 1 \leq
i \leq n \rbrace)$.

\section{Compiler and Runtime System}
\label{sec:runtime-system}

Once a survey passes the validity checks described in
Section~\ref{sec:validity-checks}, the \surveyman{} compiler
transforms it into a payload that runs inside a web page
($\S$\ref{sec:compilation}). \surveyman{} then deploys the survey,
either by acting as a webserver itself, or by posting the page to a
crowdsourcing platform and collecting responses
($\S$\ref{sec:execution}). Display of questions and flow through
the survey are managed by an interpreter written in JavaScript
($\S$\ref{sec:engine}). After all responses have been collected,
\surveyman{} performs the dynamic analyses described in
Section~\ref{sec:dynamic-analysis}.

\subsection{Compilation}
\label{sec:compilation}

\surveyman{} transforms the survey into a minimal JSON representation
wrapped in HTML, which is executed by the interpreter
described in Section~\ref{sec:engine}. The JSON representation
contains only the bare minimum information required by the JavaScript
interpreter for execution, with analysis-related information stripped
out.

Surveys can be targeted to run on different platforms; \surveyman{} supports any platform that allows the use of
arbitrary HTML. When
the platform is the local webserver, the HTML is the final
product. When posting to Amazon's Mechanical Turk (AMT), \surveyman{}
wraps the HTML inside an XML payload which is handled by
AMT. The embedded interpreter handles
randomization, the presentation of questions and answers, and
communication with the \surveyman{} runtime.

\punt{
The \surveyman{} compiler \texttt{.csv} input; the design
of the csv language is such that a human could reasonably write a
survey in it. We plan to add on a Python library that will streamline
some of the more advanced features of SurveyMan and Python proficient
users to design surveys in Python and output a JSON equivalent to a
csv survey.
}

\subsection{Survey Execution}
\label{sec:execution}

We describe how surveys appear to a respondent on Amazon's Mechanical
Turk; their appearance on a webserver is similar.  When a respondent
navigates to the webpage displaying the survey, they first see a
consent form in the HIT preview. After accepting the HIT, they begin
the survey.

The user then sees the first question and the answer options. When
they select some answer (or type in a text box), the next button and a
\textsc{Submit Early} button appear. If the question is instructional and
is not the final question in the survey, only the next button
appears. When the respondent reaches the final question, only a
\textsc{Submit} button appears.

Each user sees a different ordering of questions. However, each
particular user's questions are always presented in the same
order. The random number generator is seeded with the user's session
or \emph{assignment} id. This means that if the user navigates
away from the page and returns to the HIT, the question order will be
the same upon second viewing.

\surveyman{} displays only one question at a time. This design
decision is not purely aesthetic; it also makes measuring breakoff
more precise, because there is never any confusion as to which
question might have led someone to abandon a survey.

\subsection{Interpreter}
\label{sec:engine}

Execution of the survey is controlled by an interpreter 
 that manages the underlying evaluation and
display. This interpreter contains two key components. The first
handles survey logic; it runs in a loop, displaying questions and
processing responses. The second layer handles display information,
updating the HTML in response to certain events.

In addition to the functions that implement the state machine, the
interpreter maintains three global variables:

\begin{itemize}

\item \textbf{Block Stack:}
Since the path through a survey is determined by branching over
blocks, and since only forward branches are permitted, the interpreter
maintains blocks on a stack.

\item \textbf{Question Stack:}
Once a decision has been made as to which block is being executed
(e.g., after a random selection of a block), the appropriate questions
for that block are placed on the question stack.

\item \textbf{Branch Reference Cell:}
If a block contains a branch, this value stores its target. The interpreter
defers executing the branch until all of the questions in the block
have been answered.
\end{itemize}

The \surveyman{} interpreter is first initialized with the survey
JSON, which is parsed into an internal survey representation and then
randomized using a seeded random number generator. This randomization
preserves necessary invariants like the partial order over blocks. It
then pushes all of the top-level blocks onto the block stack. It then
pops off the first block and initializes the question stack, and
starts the survey. One question appears at a time, and the interpreter
manages control flow to branches depending on the answers given by the
respondent.

\punt{
Answers are stored in the HTML forms in the usual way. Three are three
CSS divisions (divs) in the HTML: one for the displayed question, one
for the displayed answers, and one for response input data. The
response input elements are kept in a hidden div below the answer
display div. Each new response is inserted above the previous one,
emulating a stack.
}

\punt{
The function \textsc{getNextQuestion} handles control flow and is the
primary point of interaction with the finite-state
machine. Figure~\ref{fig:getnextquestion-pseudocode} provides an
overview in pseudocode.

}

\section{Dynamic Analyses}
\label{sec:dynamic-analysis}
After collecting all results, \surveyman{} performs a series of
tests to identify survey errors and inattentive or random
respondents. Table~\ref{tab:statistical-tests} provides an overview of
the statistical tests used for each error and each possible
combination of question types tested. Different tests are required
depending on whether the questions have answers that are ordered or
unordered.

This section assumes some basic familiarity with standard
statistical tests as well as non-parametric statistical tests; for the
latter, we refer readers to Efron's classic paper on the bootstrap
method~\cite{Efron1979} and Wasserman's excellent text on
non-parametric statistics~\cite{Wasserman:2006:NS:1202956}.

\begin{table*}[!t]
\centering
\begin{tabular}{>{\small}l|>{\small}l|>{\small}l|>{\small}l}
\textsc{Error}        & \textsc{Both Ordered} & \textsc{Both Unordered} & \textsc{Ordered-Unordered} \\
\hline
\textbf{Correlated Questions}  &  Spearman's $\rho$       & Cramer's $V$ & Cramer's $V$ \\
\textbf{Question Order Bias}   &  Mann-Whitney U-Test     & $\chi^2$     &  N/A \\
\textbf{Question Wording Variant Bias} &  Mann-Whitney U-Test     & $\chi^2$     &  N/A \\
\textbf{Inattentive or Random Respondents} & Nonparametric bootstrap & Nonparametric bootstrap &  N/A \\
                               & over empirical entropy    & over empirical entropy  &  N/A \\
\end{tabular}
\caption{The statistical tests used to find particular errors. Tests are conducted pair-wise across questions; each column indicates whether the pairs of questions both have answers that are ordered, both unordered, or a mix.\label{tab:statistical-tests}}
\end{table*}

\subsection{Correlated Questions}
\label{sec:analysis-correlated-questions}

\surveyman{} analyzes correlation in two ways. The \textsc{Correlated}
column can be used to indicate sets of questions that the survey author
expects to have statistical correlation. Flagged questions can be used
to validate or reject hypotheses and to help detect bad
actors. Alternatively, if a question not marked as correlated
is found to have a statistically significant correlation, then \surveyman{} flags it.

Questions are compared pair-wise.
\surveyman{} supports automated correlation analysis only between
exclusive (radio button) questions. These questions may be ordered or
unordered.

For two questions such that at least one of them is unordered,
\surveyman{} returns the $\chi^2$ statistic, its $p$-value, and
computes Cramer's $V$ to determine correlation. It also uses Cramer's
$V$ when comparing an unordered and an ordered
question. Ordered questions are compared using Spearman's $\rho$.  In
practice, \surveyman{} rarely has sufficient data to return confidence
intervals on such point estimates. Instead, it simply flags the pair
of questions.

This survey author can act on this information in two
ways. First, the survey author may decide to shorten the survey by removing
one or more of the correlated questions. It is ultimately the
responsibility of the survey author to use good judgement and domain
knowledge when deciding to remove questions. Second, the survey author
could use discovered correlations to assist in identification of
cohorts or bad actors by updating the entries in the
\textsc{Correlated} column appropriately.

\subsection{Question Order Bias}
\label{sec:analysis-question-order-bias}

To compute order bias, \surveyman{} uses the Mann-Whitney U test for
ordered questions and the $\chi^2$ statistic for unordered
questions. In each case, \surveyman{} attempts to rule out the null
hypothesis that the distributions of responses are the same regardless
of question order.

For each question pair $(q_i, q_j)$, where $i \neq
j$, \surveyman{} partitions the sample into two sets: $S_{i < j}$,
the set of questions where $q_i$ precedes $q_j$, and $S_{j < i}$, the
set of questions where $q_i$ follows $q_j$. \surveyman{} assumes each
set is independent. We outline below how to test for bias in $q_i$
when $q_j$ precedes it (the other case is symmetric), both for ordered
and unordered questions.

\paragraph{Ordered Questions: Mann-Whitney $U$ Statistic}

\begin{enumerate}

\item Assign ranks to each of the options. For example, in a Likert-scale question having options \emph{Strongly Disagree}, \emph{Disagree}, \emph{Agree}, and \emph{Strongly Agree}, assign each the values 1 through 4.

\item Convert each answer to $q_i$ in $S_{i<j}$ to its rank, assigning average ranks to ties.

\item Convert each answer to $q_j$ in $S_{j<i}$ to its rank, assigning average ranks to ties.

\item Compute the $U$ statistic over the two sets of ranks. If the probability of computing $U$ is less than the critical value, there is a significant difference in the ordering.

\end{enumerate}

\paragraph{Unordered Questions: $\chi^2$ Statistic}
\begin{enumerate}

\item Compute frequencies $f_{i < j}$ for the answer options of $q_i$ in the set of responses $S_{i<j}$. We use these values to compute the estimator.

\item Compute frequencies $f_{j < i}$ for answer options $q_i$ in the set of responses $S_{j < i}.$ These form our observations.

\item Compute the $\chi^2$ statistic on the data set. The degrees of
freedom will be one less than the number of answer options,
squared. If the probability of computing such a number is less than
the value at the $\chi^2$ distribution with these parameters, there is
a significant difference in the ordering.

\end{enumerate}

\surveyman{} computes these values for every unique question pair, and
reports questions with an identified order bias.

\subsection{Question Wording Variant Bias}
\label{sec:analysis-wording-variant-bias}

Wording bias uses almost the same analysis approach as order bias. 
Instead of comparing two sets of responses, \surveyman{} compares $k$
sets of responses, where $k$ corresponds to the number of variants. As
with order bias, \surveyman{} reports questions whose wording variants
lead to a statistically significant difference in responses.

\subsection{Breakoff vs. Fatigue}
\label{sec:analysis-breakoff}

\surveyman{} identifies and distinguishes two kinds of breakoff:
breakoff triggered at a particular \emph{position} in the survey, and
breakoff at a particular \emph{question}.  Breakoff by position is
often an indicator that the survey is too long. Breakoff by question
may indicate that a question is unclear, offensive, or burdensome to
the respondent.

Since \surveyman{} randomizes the order of questions
whenever possible, it can generally distinguish between positional
breakoff and question breakoff without the need for any statistical
tests. To identify both forms of breakoff, \surveyman{} reports ranked
lists of the number of respondents who abandoned the survey by
position and by question. A cluster around a position indicates
fatigue or that the compensation for the survey, if any, is
inadequate. A high number of abandonments at a particular question
indicates a problem with the question itself.

\subsection{Inattentive or Random Respondents}
\label{sec:analysis-random-respondents}

When questions are unordered, inattentive respondents (who, for
example, always click on the first choice) are indistinguishable from
random respondents. The same analysis thus identifies both types of
respondents automatically. This analysis is most effective when the
survey itself has a reasonable maximum entropy, which \surveyman{}
computes in its static analysis phase
($\S$\ref{sec:static-analysis-reports}).

The longer a survey is and the more question options it has, the
greater the dimensionality of the space. Random respondents will
exhibit higher entropy than non-random respondents, because they will
uniformly fill the space. We thus employ an entropy-based test to
identify them.

\surveyman{} first removes all respondents who did not complete the survey, since
their entropy is artificially low and would appear to be non-random
respondents.
\surveyman{} then computes the empirical probabilities for each
question's answer options. Then for every response $r$, it calculates a
score based on entropy: $\mbox{score}_r = \sum_{i=1}^n
p(o_{r,q_i})∗\log_2(p(o_{r,q_i}))$.
\surveyman{} uses the bootstrap method to define a
one-sided 95\% confidence interval. It then reports any respondents
whose score falls outside this interval.

\section{Evaluation}
\label{sec:evaluation}

We evaluate \surveyman{}'s usefulness in a series of case studies with
surveys produced by our social scientist colleagues. We address the
following research questions:

\begin{itemize}

\item \textbf{Research Question 1:} Is \surveyman{} usable by survey authors and sufficiently expressive to describe their surveys?
\item \textbf{Research Question 2:} Is \surveyman{} able to identify survey errors?
\item \textbf{Research Question 3:} Is \surveyman{} able to identify random or inattentive respondents?
\end{itemize}

\begin{figure}[!t]
 \centering
 \begin{tabular}{>{\small}l>{\small}l}
 \hline
 \multicolumn{2}{l}{Please say both words aloud. Which one would you say?} \\
 $\bigcirc$ & definitely antidote-athon \\
 $\bigcirc$ & probably antidote-athon \\
 $\bigcirc$ & probably antidote-thon \\
 $\bigcirc$ & definitely antidote-thon \\
 \end{tabular}
  \caption{An example question used in the phonology case study ($\S$\ref{sec:case-study-phonology}).\label{fig:alicious}
}
\end{figure}

\subsection{Case Study 1: Phonology}
\label{sec:case-study-phonology}

The first case study is a phonological survey that tests the rules by
which English-speakers are believed to form certain word
combinations. This survey was written in \surveyman{} by a colleague
in the Linguistics department at the University of Massachusetts with
limited guidance from us (essentially an abbreviated version of
Section~\ref{sec:language}).

The first block asks demographic questions, including age and whether
the respondent is a native speaker of English. The second block
contains 96 Likert-scale questions. The final block consists of one
freetext question, asking the respondent to provide any feedback they
might have about the survey.

Each of the 96 questions in the second block asks the respondent to
read aloud an English word suffixed with either of the pairs
``-thon/-athon'' or ``licious/-alicious'' and judge which sounds more
like an English word. An example appears in Figure~\ref{fig:alicious}.

Our colleague first ran this survey in a controlled experiment
(in-person, without \surveyman{}) that provides a gold-standard data
set. We ran this survey four times on Amazon's Mechanical Turk between
September 2013 and March 2014 to test our techniques.

\paragraph{Static Analysis:}
This survey has a maximum entropy of 195.32; the core
96 questions have a maximum entropy of 192 bits. There is no
branching in this survey, so without breakoff, the minimum, maximum,
and average path lengths are all 99 questions long.

\paragraph{Dynamic Analysis:}
The first run of the survey was early in \surveyman{}'s development
and functioned primarily as a proof of concept. There was no quality
control in place. We sent the results of this survey to our
colleagues, who verfied that random respondents were a major issue and
were tainting the results, demonstrating the need for quality control.

The latter three runs were performed at different times of day under
slightly different conditions (e.g., time of day); all three permitted
breakoff. \surveyman{} detected random respondents in all three
cases. \surveyman{} found approximately 6\% of the respondents to the
first two surveys were inattentive or random. We used minimal
Mechanical Turk qualifications (at least one HIT done with an 80\% or
higher approval rate) to filter the first survey, so we expected to
see fewer random respondents. Somewhat surprisingly, we found that
qualifications made no difference: the second run, launched in the
morning on a work day without qualifications, produced similar results
to the first. The third run was launched on a weekend night and
approximately 15\% of respondents were identified as responding
randomly. Upon looking closely at these respondents, we saw strong
positional preferences (i.e., respondents were repeatedly clicking the
same answer), corroborating \surveyman{}'s findings.

Because the survey lacks structure---effectively it consists of only
one large block, with no branching---and consists of roughly uniform
questions, we did not expect to find any biases or other errors,
and \surveyman{} did not report any.


\subsection{Case Study 2: Psycholinguistics}
\label{sec:case-study-psycholinguistics}

\begin{figure}[!t]
\centering
\begin{tabular}{>{\small}l>{\small}l}
\hline
\multicolumn{2}{l}{How odd is the number 3?} \\
$\bigcirc$ & Not very odd. \\
$\bigcirc$ & Somewhat not odd. \\
$\bigcirc$ & Somewhat odd. \\
$\bigcirc$ & Very odd. \\
\end{tabular}
\caption{An example question used in the psycholinguistics case study ($\S$\ref{sec:case-study-psycholinguistics}).}\label{fig:parity}
\end{figure}

The second case study is a test of what psychologists
call \emph{prototypicality} and was written in the \surveyman{}
language by another one of our colleagues in the Linguistics
department; as with our other colleague, she wrote this survey with
only minimal guidance from us.

In this survey, respondents are asked to use a scale to rank how well
a number represents its parity (e.g., ``how odd is this
number?''). The goal is to test how people respond to a categorical
question (since numbers are only either even or odd) when given a
range of subjective responses.

There are 65 questions in total. The survey is composed of two
blocks. The first block is a floating block that contains 16 floating
subblocks of type \textsc{All}; that is, only one randomly-chosen
question from the possible variants will be asked. Every question in
one of these floating blocks has slightly different wording for both
question and options. The other block contains one question asking the
respondent about their native language. Because the first block is
floating, this question can appear either at the beginning or the end
of the survey.

We launched this survey on a weekday morning. It took about a day to
reach our requested level of 150 respondents.

\paragraph{Static Analysis:}
The maximum entropy for the survey is 34 bits. Since there is no true
branching in the survey, every respondent sees the same number of
questions, unless they submit early. The maximum, minimum, and average
survey lengths are all 17 questions.

\paragraph{Dynamic Analysis:}

\surveyman{} found no significant breakoff or order bias, but it did
find that several questions exhibited marked wording variant
bias. These were for the numbers 463, 158, 2. For the number 463,
these pairs had significantly different responses:

\begin{enumerate}

\item \emph{How well does the number 463 represent the category of odd numbers?} and \emph{How odd is the number 463?} 

\item \emph{How well does the number 463 represent the category of odd numbers?} and \emph{How good an example of an odd number is the number 463?}

\item \emph{How odd is the number 463?} and \emph{How good an example of an odd number is the number 463?}
\end{enumerate}

While these questions may have appeared to the survey author to be
semantically identical, they in fact lead to substantial differences
in the responses. This result confirms our colleagues' hypothesis,
which is that previous studies are flawed because they do not take
question variant wording bias into account.

\subsection{Case Study 3: Labor Economics}
\label{sec:case-study-labor}
Our final case study is a survey about wage negotiation, conducted
with a colleague in Labor Studies. The original design was completely
flat with no blocks or branching. There were 38 questions that were a
mix of Likert scales, checkbox, and other unordered questions. The survey
asked for demographic information, work history, and attitudes about
wage negotiation.

Our collaborator was interested in seeing whether complete
randomization would have an impact on the quality of results. We
worked with her to write a version of the survey that comprised both a
completely randomized version and a completely static version: a
respondent is randomly given one of the two versions. This approach
lets us to collect data under nearly identical conditions, since each
respondent is equally likely to see the randomized or the static
version.

We launched this survey on a Saturday and obtained 69 respondents.  We
observed extreme survey abandonment: none of the respondents completed
the survey. The maximum number of questions answered was 26.

\paragraph{Static Analysis:}
\surveyman{} reports that this survey has a maximum entropy of 342
bits, which is extremely high; some questions have numerous options,
such as year of birth. Every path is of length 40, as expected.

\paragraph{Dynamic Analysis:}
The analysis of breakoff revealed both positional and question
breakoff. 72.5\% of the breakoff occurred in the first six positions,
suggesting that the compensation for the survey was insufficient
(\$0.10 USD). 52\% of the instances of breakoff also occurred in just five of
the questions, shown in Table~\ref{fig:breakoff_question}.

These results illustrate the impact of randomization on diagnosing
breakoff. If question order were fixed for the whole population, we
would not be able to determine whether question or position was
causing breakoff. In the ordered version, the seven questions are all
demographic. It is reasonable to conclude that some respondents find
these demographic questions to be problematic, leading them to abandon
the survey. This suggests that the survey author should consider
placing these questions at the end of the survey, or omitting them if
possible.


\begin{figure*}[!t]
\centering
\begin{tabular}{>{\small}l>{\small}l>{\small}l}
\hline
\textbf{Question} & \textbf{Count} & \textbf{Version}\\ \hline
Please choose one.&12&All\\
In which country were you born?&12&Static\\
In what year were you born?&5&Static\\
In which country do you live in now?&4&Static\\
In the past, have you ever asked your manager or boss for an increase in pay or higher wage?&3&Fully Randomized\\
\end{tabular}
\caption{The top five question breakoffs found in the psycholinguistics case study ($\S$\ref{sec:case-study-labor}).}\label{fig:breakoff_question}
\end{figure*}

\section{Related Work}
\label{sec:related-work}
\begin{table*}[!t]
\centering
\begin{tabular}{>{\small}l|>{\small}l|>{\small}c|>{\small}c|>{\small}c|>{\small}c}
\hline
\textsc{Language} & \textsc{Type} & \textsc{Loops} & \textsc{Question} & \textsc{Error} & \textsc{Random} \\
 &  &  & \textsc{Randomization} & \textsc{ Detection} & \textsc{Respondent Detection} \\
\hline
\textbf{Blaise}~\cite{blaise} & Standalone DSL & & & & \\
\textbf{QPL}~\cite{qpl}       & Standalone DSL & & & & \\
\textbf{Topsl}~\cite{MacHenry-Matthews:Scheme04} & Embedded DSL & \cmark & & & \\
\textbf{websperiment}~\cite{mackerron2011} & Embedded DSL & \cmark & & & \\
\textbf{SuML}~\cite{barclay2002suml}      & XML schema & & & & \\
\textbf{SQBL}~\cite{sqbl}     & XML schema & \cmark & & & \\
\hline
\textbf{\textsc{SurveyMan}} & Standalone DSL & & \cmark & \cmark & \cmark \\
\hline
\end{tabular}
\caption{Feature comparison of previous survey languages and \surveyman{}.\label{tab:features}}
\end{table*}


\subsection{Survey Languages}

Table~\ref{tab:features} provides an overview of previous survey
languages and constrasts them with \surveyman{}.

\paragraph{Standalone DSLs.}
Blaise is a language for designing survey logic and
layout~\cite{blaise}. Blaise programs consist of named blocks that
contain question text, unique identifiers, and response type
information. Control flow is specified in a rules block, where users
list the questions by unique identifier in their desired display
order. This rules block may also contain data validation rules, such
that an Age field be greater than 18. 
Similarly, QPL is a language and deployment tool for web surveys
sponsored by the U.S. Government Accountability
Office~\cite{qpl}. Neither language supports randomization or identifies survey errors.

\paragraph{Embedded DSLs.}
Topsl~\cite{MacHenry-Matthews:Scheme04} and
websperiment~\cite{mackerron2011}  embed survey structure
specifications in general-purpose programming languages (PLT Scheme
and Ruby, respectively). Topsl is a library of macros 
to express the content, control flow, and layout of
surveys; Websperiment provides similar functionality.
Both provide only logical structures
for surveys and a means for customizing their presentation. Neither one
can detect survey errors or do quality control.

\paragraph{XML-Based Specifications.}
At least two attempts have been made to express survey structure using
XML. The first is SuML; the documentation for this schema is no longer
available and its specification is not described in its
paper~\cite{barclay2002suml}. A recent XML implementation of survey
design is SQBL~\cite{sqbl,spencer2013}. SQBL is available with a
WYSIWYG editor for designing surveys~\cite{canard}. It offers default
answer options, looping structures, and reusable components. It is the
only survey specification language we are aware of that is extensible,
open source, and has an active community of users.
Unlike \surveyman{}, none of these provide for randomization or
error analyses.



\subsection{Web Survey Tools}

There are numerous tools for deploying web-based surveys, including
Qualtrics, SurveyMonkey, instant.ly, and Google Consumer
Surveys~\cite{qualtrics,surveymonkey,instantly,googlesurveys}. Because these
survey tools impose a strict order on questions and offer no way to
express variants, they cannot control for question ordering or
question choice bias. Some of these systems, like Google Consumer
Surveys, offer the ability to randomize the order of
\emph{answers}, but not \emph{questions}~\cite{googlesurveys}. None of these systems address the problem of debugging survey errors or inattentive respondents.

\subsection{Survey Analyses}
\label{sec:automatedqc}

Despite the fact that surveys are a well-studied topic---a key survey
text from 1978 has over 11,000
citations~\cite{dillman1978mail})---there has been surprisingly little
work on approaches to automatically address survey errors.
We are aware of no previous work on identifying any other errors
beyond breakoff. Peytchev et al.\ observe that there is little
scholarly work on identifying breakoff and attribute this to a lack of
variation in question characteristics or randomization of question
order, which \surveyman{} provides~\cite{peytchev2009survey}.

\paragraph{Inattentive Respondents.}
Several researchers have proposed \emph{post hoc} analyses to identify
inattentive respondents, also known as \emph{insufficent effort
responding}~\cite{huang2012detecting}. Meade et al. describe a
technique based on multivariate outlier analysis (Mahalanobis
distance) that identifies
respondents who are consistently far from the mean of a set of items~\cite{meade2012identifying}.

Unfortunately, this technique assumes that legitimate survey responses
will all form one cluster. This assumption can easily be violated if
survey respondents constitute multiple clusters. For example,
politically conservative individuals might answer questions on a
health care survey quite differently from politically liberal
individuals. In any event, there is no compelling reason to believe
clusters will be distributed normally around the mean.

Other \emph{ad hoc} techniques aimed at identifying random or
low-effort respondents include measuring the amount of time spent
completing a survey, adding CAPTCHAs, and measuring the entropy of
responses, assuming that low-effort respondents either choose one
option or alternate between options
regularly~\cite{zhucarterette2010}; none of these reliably
distinguish lazy or random respondents from real respondents.



\subsection{Randomized Control Flow}
As far as we are aware, \surveyman{}'s language is the first to
combine branches with randomized control flow. Dijkstra's Guarded
Command notation requires non-deterministic choice among any true
cases in
conditionals~\cite{Dijkstra:1975:GCN:360933.360975}. It should come as no surprise that
Dijkstra's language does not support \textsc{Goto} statements;
\surveyman{} supports branches from and to randomized
blocks.

\subsection{Crowdsourcing and Quality Control}
\label{sec:crowdsourcing}

Barowy et al.\ describe \textsc{AutoMan}, an embedded domain-specific
language that integrates digital and human computation via
crowdsourcing platforms~\cite{barowy2012}. Both \textsc{AutoMan} and
\surveyman{} have shared goals, including automatically ensuring
quality control of respondents. However, \textsc{AutoMan}'s focus is
on obtaining a single correct result for each human
computation. \surveyman{} instead collects \emph{distributions} of responses, which requires an entirely different approach.


\section{Conclusion and Future Work}
\label{sec:conclusion}
This paper reframes surveys and their errors as a programming language
problem. It presents \surveyman{}, a programming language and runtime
system for implementing surveys and identifying survey errors and
inattentive respondents. Pervasive randomization prevents small biases
from being magnified and enables statistical analyses, letting
\surveyman{} identify serious flaws in survey design. We
believe that this research direction has the potential to have
significant impact on the reliability and reproducibility of research
conducted with surveys.
\surveyman{} is available for download at \url{http://www.surveyman.org}.

\acks

This material is based upon work supported by the National Science
Foundation under Grant No. CCF-1144520. The authors would like to
thank our social science collaborators at the University of
Massachusetts, Sara Kingsley, Joe Pater, Presley Pizzo, and Brian Smith; John
Foley, Molly McMahon, Alex Passos, and Dan Stubbs; and fellow PLASMA
lab members Dan Barowy, Charlie Curtsinger, and John Vilk for valuable
discussions during the evolution of this project.

\punt{
}
{
\bibliographystyle{abbrv}
\bibliography{ref}
}

\end{document}